\documentclass[aps,prl,reprint,superscriptaddress,nofootinbib,floatfix]{revtex4-2}

\usepackage[T1]{fontenc}
\usepackage[utf8]{inputenc}
\usepackage{graphicx}
\usepackage{amsmath,amssymb,bm}
\usepackage{microtype}
\usepackage[colorlinks=true,allcolors=blue]{hyperref}
\begin{document}

\title{Exponential Control of Adiabatic Time Reflection by Complex-Time Singularities}

\author{Shiyue Deng}
\affiliation{International Center for Quantum Design of Functional Materials (ICQD), Hefei National Research Center for Interdisciplinary Sciences at the Microscale, University of Science and Technology of China, Hefei, Anhui 230026, China}
\affiliation{Hefei National Laboratory, University of Science and Technology of China, Hefei 230088, China}
\author{Qian Niu}
\affiliation{International Center for Quantum Design of Functional Materials (ICQD), Hefei National Research Center for Interdisciplinary Sciences at the Microscale, University of Science and Technology of China, Hefei, Anhui 230026, China}
\affiliation{School of Physics, University of Science and Technology of China, Hefei, Anhui 230026, China}
\date{\today}

\begin{abstract}
Nearly identical temporal switching profiles can produce very different reflections even when both have small local adiabatic measures. In a lossless dielectric, we obtain this difference by adding a weak feature that changes the complex-time singularities while preserving monotonicity, endpoints, and central slope. A perturbation bounded by $0.6\%$ of the total inverse-permittivity change raises the reflected mode amplitude from $8.02\times10^{-10}$ to $1.28\times10^{-3}$. The complex action and pole residue predict this amplitude to within $0.32\%$. Within this waveform family, deviation and monotonicity constraints select three design regimes. At fixed deviation bound, reoptimization keeps the central feature's time scale comparable to the inverse incident angular frequency. The optimal amplitude is then proportional to $1/T$ as the switching time $T$ tends to infinity. Stretching a fixed profile gives exponential suppression. Full-wave Maxwell simulations resolve the backward packet, and filtering calculations quantify its sensitivity to modulation bandwidth.
\end{abstract}

\maketitle
\raggedbottom

\begin{figure*}[t]
\centering
\includegraphics[width=\textwidth]{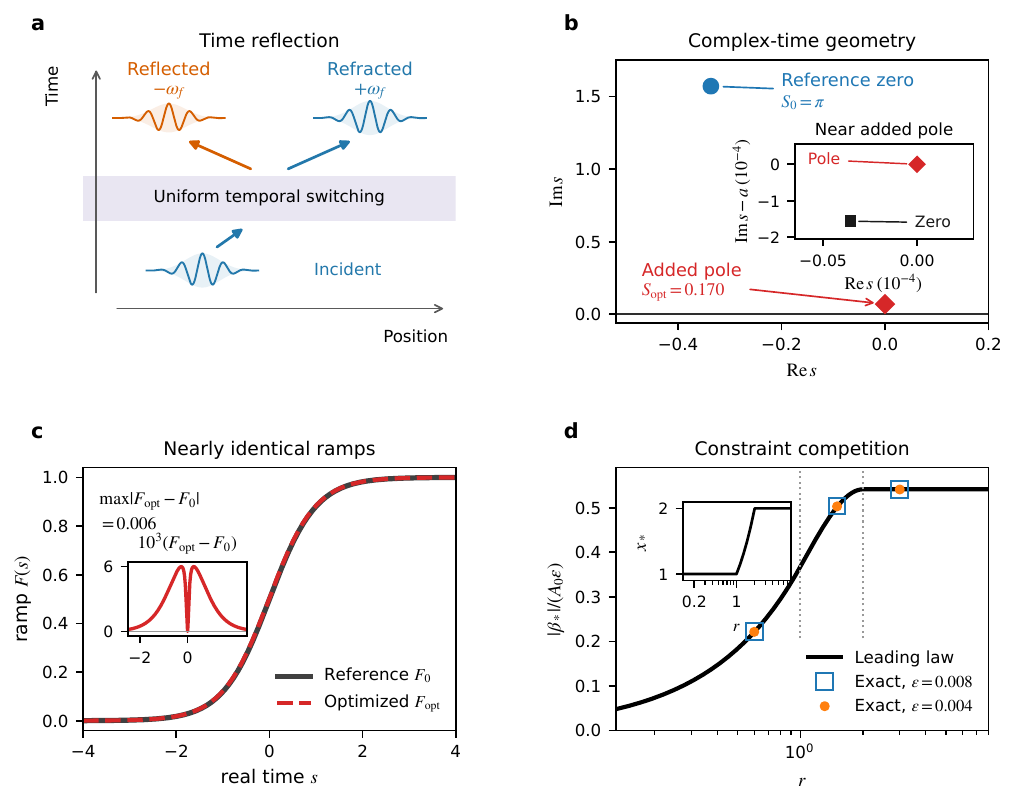}
\caption{\label{fig:design}Physical picture and constrained complex-time design. (a) A spatially uniform temporal switch produces forward and backward packets. The horizontal band represents switching everywhere in space; arrows indicate propagation as time advances. The outgoing branches have frequencies $+\omega_f$ and $-\omega_f$ at conserved wave vector. Packet shapes and trajectories are schematic, and the reflected packet is enlarged. (b) The reference turning point and engineered pole. The inset resolves the pole (red diamond) and adjacent zero (black square) of $\Omega_\eta^2$, using coordinates relative to $ia$ in units of $10^{-4}$. The actions are $S=0.1703$ and $S_0=\pi$. (c) Reference and selected ramps, $F_0$ and $F_{\rm opt}$, at $\epsilon=0.15$, differing by at most $0.006$ of the full range. (d) Optimal scaled reflection amplitude versus constraint ratio $r$: the curve is Eq.~\eqref{eq:constraintLaw}, and symbols are full mode amplitudes at designs selected by complex-time theory. The inset shows the optimal scaled feature width $x_*$; quantities are defined around Eq.~\eqref{eq:constraintLaw}.}
\end{figure*}

\begin{figure*}[t]
\centering
\includegraphics[width=\textwidth]{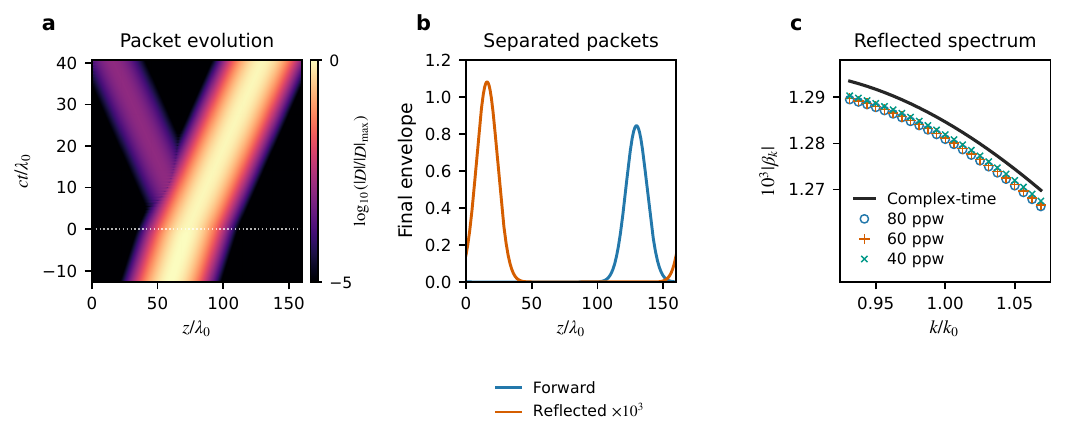}
\caption{\label{fig:fullwave}Full-wave wave-packet realization. (a) Spacetime field map after the selected spatially uniform temporal modulation. The negative-frequency component propagates backward. (b) Separated forward and reflected packets; the reflected envelope is rescaled for visibility. (c) Reflected amplitude over the resolved packet bandwidth. The curve is the complex-time prediction; symbols are full-wave results at three resolutions in points per initial carrier wavelength (ppw). Here $z$ is position, $k_0$ the central wave number, $\lambda_0=2\pi/k_0$, and $D$ the displacement field normalized to the initial peak.}
\end{figure*}

Temporal modulation controls the frequency and propagation of electromagnetic waves~\cite{galiffi2022review,morgenthaler1958,Deng2026}. An abrupt change in a spatially homogeneous medium couples positive- and negative-frequency waves at conserved momentum, producing frequency translation and a backward, phase-conjugated field~\cite{xiao2014}. Electromagnetic time reflection has been observed in switched transmission lines and optically controlled microwave platforms~\cite{moussa2023,jones2024}. Real interfaces switch over a finite time. Wentzel--Kramers--Brillouin (WKB) theory describes soft temporal boundaries~\cite{hadad2020}. Temporal tapers and time-domain supersymmetry can suppress reflection or prescribe a real-frequency response~\cite{galiffi2022taper,silbiger2023,antyufeyeva2025,hou2026,garciameca2020}.

For the fixed analytic profiles considered here, slow switching exponentially suppresses time reflection. These profiles have constant asymptotic endpoints and a nonzero mode frequency, and their outgoing backward amplitude is smaller than every power of the inverse switching time. We compare nearly identical real-time profiles with small local adiabatic measures to determine how much their reflections can differ.

Dykhne related weak adiabatic transitions to complex-time degeneracies~\cite{dykhne1962}; Dingle developed large-order asymptotics~\cite{dingle1973}. Berry connected analytic continuation to weak above-barrier reflection~\cite{berry1982} and established Stokes smoothing under optimal truncation~\cite{berry1989}. The same Stokes structure governs positive--negative-frequency mixing in time-dependent oscillators~\cite{dumlu2010,dabrowski2016}. Related enhancement occurs in dynamically assisted Schwinger production, where a weak fast field enhances an exponentially suppressed process~\cite{schutzhold2008}. Complex turning points also determine its pulse-shape-dependent interference~\cite{dumlu2011}. We determine the enhancement allowed by monotonicity, fixed endpoints and central slope, and a maximum waveform deviation of $0.006$. These constraints select three optimal-width regimes and an algebraic reflection envelope.

\section*{Model and exponential reflection}

We consider a homogeneous, lossless dielectric with vacuum permeability $\mu_0$ and relative permittivity $\varepsilon_r(t)$ varying with time $t$. Spatial uniformity conserves the wave vector, of magnitude $k$. For either transverse polarization, the displacement-field Fourier amplitude $D_k$ obeys
\begin{equation}
\begin{gathered}
\epsilon^2D_k''(s)+\Omega_k^2(s)D_k(s)=0,\\
s=t/T,\qquad \epsilon=(\omega_0T)^{-1},\\
\Omega_k(s)=ck/[\omega_0\sqrt{\varepsilon_r(s)}].
\end{gathered}
\label{eq:oscillator}
\end{equation}
Here $c$ is the vacuum light speed, $T$ the switching time, $s$ dimensionless time, and a prime denotes $d/ds$. The dimensionless frequency is $\Omega_k=\omega_k/\omega_0$, with initial and final values $\Omega_i,\Omega_f$. We choose $\omega_0=\omega_i(k)=ck/\sqrt{\varepsilon_{r,i}}$, where $\varepsilon_{r,i}$ is the initial permittivity, so $\Omega_i=1$ and $\epsilon\ll1$ measures adiabaticity. A unit positive-frequency input has the output
\begin{equation}
D_k^{\mathrm{out}}=
\alpha_k\frac{e^{-i\Omega_fs/\epsilon}}{\sqrt{2\Omega_f}}
+\beta_k\frac{e^{+i\Omega_fs/\epsilon}}{\sqrt{2\Omega_f}},
\qquad |\alpha_k|^2-|\beta_k|^2=1.
\label{eq:bogoliubov}
\end{equation}
Here $\alpha_k$ and $\beta_k$ are the time-refraction and time-reflection amplitudes. At conserved $k$, the negative-frequency branch has opposite group velocity and therefore propagates backward, as sketched in Fig.~\ref{fig:design}(a).

In an instantaneous basis, the changing medium couples positive- and negative-frequency modes. Contributions generated at different times largely cancel during slow switching. For our fixed analytic ramps, the final reflection is exponentially small, as the exact tanh solution illustrates. The negative-frequency coefficient can be much larger during switching, before later contributions cancel the earlier mixing~\cite{supp}.

Finite-order local WKB corrects the positive-frequency solution but misses this final reflection. Complex-time analysis describes its appearance through the Stokes phenomenon in the analytic continuation of $\Omega_k^2(s)$. In our examples, the relevant complex feature $s_c$ is a zero or pole. At a zero the frequency branches meet and local WKB fails; a pole makes the mode equation singular. Both make the mode-coupling coefficient $\Omega_k'/(2\Omega_k)=(\Omega_k^2)'/(4\Omega_k^2)$ singular, but they have different connection laws. When one contribution dominates,
\begin{equation}
\begin{aligned}
|\beta_k|&\sim A_k\epsilon^{-p}e^{-S_k/\epsilon},\\
S_k&=2\left|\operatorname{Im}\int_{s_*}^{s_c}\Omega_k(s)\,ds\right|,
\qquad s_*\in\mathbb R.
\end{aligned}
\label{eq:generalCT}
\end{equation}
Moving the real reference point $s_*$ changes only the real part of the action. The frequency is continued from its positive real-time value along the chosen contour. The exponent $p$ and prefactor $A_k$ depend on the feature's local type and strength.

These complex features lie away from real time. Ramps that are close on the real axis can have very different complex actions and final reflections.

Suppressing the fixed-mode index $k$, we test this description against the solvable reference ramp $F_0$,
\begin{equation}
F_0(s)=\frac{1+\tanh s}{2},\qquad
\Omega^2(s)=\Omega_i^2+(\Omega_f^2-\Omega_i^2)F_0(s).
\label{eq:tanh}
\end{equation}
Since $\Omega^2\propto1/\varepsilon_r$ at fixed $k$, this is a tanh profile of the inverse permittivity between its initial and final values. The hypergeometric solution~\cite{alves2019} gives the reference reflection amplitude $\beta_0$:
\begin{equation*}
|\beta_0|^2=
\frac{\sinh^2[\pi(\Omega_f-\Omega_i)/(2\epsilon)]}
{\sinh(\pi\Omega_i/\epsilon)\sinh(\pi\Omega_f/\epsilon)}.
\end{equation*}
For the endpoint values $\Omega_i=1$, $\Omega_f=1.4$ used below, its adiabatic tail is $|\beta_0|\sim e^{-\pi/\epsilon}$, consistent with the action $S_0=\pi$ of the nearest complex turning point. The Supplemental Material tests large-order growth and optimal truncation~\cite{supp}.

\section*{Complex-time design}

We add a weak feature that changes the analytic continuation while keeping the real-time ramp nearly fixed:
\begin{equation}
\begin{aligned}
F_\eta(s)&=F_0(s)+\eta\frac{s^2}{s^2+a^2}\operatorname{sech}^2s,\\
\Omega^2(s)&=\Omega_i^2+(\Omega_f^2-\Omega_i^2)F_\eta(s).
\end{aligned}
\label{eq:family}
\end{equation}
Here $\eta\geq0$ is the perturbation strength, $a>0$ its dimensionless width, and $\operatorname{sech}s=1/\cosh s$. The added term preserves the endpoints, midpoint, and central slope of $F_0$. Its poles at $s=\pm ia$ are inherited by $\Omega_\eta^2$ through Eq.~\eqref{eq:family}. For $a\ll1$, the central feature varies on the shorter time scale $aT$, while its envelope varies on the overall switching scale $T$.

At $\eta=0$, the derivative of the reflection amplitude with respect to perturbation strength has an exact integral expression in terms of the reference solutions. Approximating these solutions by WKB and evaluating the residue at $s=-ia$ gives a complex-time prediction for the modified amplitude $\beta_\eta$ (derivation in the Supplemental Material~\cite{supp}):
\begin{equation}
|\beta_\eta-\beta_0|\sim
\eta\frac{C(a)}{\epsilon}e^{-S(a)/\epsilon},
\qquad
S(a)=2\operatorname{Im}\int_0^{ia}\Omega_0(s)\,ds,
\label{eq:residue}
\end{equation}
Here $\Omega_0$ is the reference frequency. We define $\Delta_\Omega^2\equiv\Omega_f^2-\Omega_i^2$ as the total change in the squared frequency. The residue prefactor is $C(a)=\pi\Delta_\Omega^2a\sec^2a/[2|\Omega_0(-ia)|]$. The factor $1/\epsilon$ comes from linear scattering; $S$ is the background action. The modified $\Omega_\eta^2(s)$ also has a zero close to the added pole [Fig.~\ref{fig:design}(b)]. As the perturbation vanishes, the zero approaches the pole and the added structure disappears. The rule accounts for their combined first-order effect when background WKB is valid and the pole contribution dominates.

At $\epsilon=0.15$, we optimize $a$ and $\eta$ using Eq.~\eqref{eq:residue}, subject to $F_\eta'\geq0$ and $\max|F_\eta-F_0|\leq0.006$. Monotonicity retains a single smooth switch without material-parameter reversals. The maximum deviation is $0.6\%$ of $F$'s total change. The selected near optimum (subscript $\mathrm{opt}$) has $a=0.070$, $\eta=0.00688$, and $S=0.1703$. The local adiabatic measure $\mathcal A(s)=\epsilon|\Omega'|/\Omega^2$ measures the fractional frequency change over a time interval $1/\omega(t)$. Its peak changes only from $0.0213$ to $0.0221$.

Numerical integration of the full oscillator equation gives $|\beta_{\rm opt}|=1.28055\times10^{-3}$, compared with $|\beta_0|=8.01785\times10^{-10}$. Equation~\eqref{eq:residue} predicts $1.28464\times10^{-3}$, a difference of $0.32\%$. We compare $\Delta\beta=\beta_\eta-\beta_0$ in common asymptotic phases and retain the reference contribution whenever it is appreciable. Replacing the envelope by $e^{-2s^2}$ changes the residue but preserves the pole position and background action. For both envelopes, predictions agree with full mode integration within $0.58\%$ at three widths. The Supplemental Material separates the approximation errors~\cite{supp}.

The relative phase between positive- and negative-frequency modes accumulates at a rate $2\omega(t)$, which gives a spectral interpretation~\cite{galiffi2022taper}. As switching slows, the reference derivative $dF_0/dt$ has exponentially little spectral weight near these frequencies. The narrow feature adds weight there while keeping the local adiabatic measure small. Its duration is $aT\simeq0.47/\omega_0$. At this design, $S/\epsilon=1.14$, so the small reflected amplitude depends on both exponential suppression and weak perturbation strength.

For $a\ll1$, the pole at $ia$ approaches the switching center $s=0$.
Taking $\Omega_{\rm mid}=\Omega_0(0)$ along the short contour reduces Eq.~\eqref{eq:residue} to
\begin{equation}
 |\Delta\beta|\simeq
 \frac{\eta\pi\Delta_\Omega^2}{2\Omega_{\rm mid}}
 \frac{a}{\epsilon}e^{-2\Omega_{\rm mid}a/\epsilon}.
 \label{eq:narrowPole}
\end{equation}
Bringing the pole closer reduces exponential suppression but also reduces
the residue, $C(a)\propto a$. These competing effects select a finite
optimal width even at fixed strength.

Let $\delta$ bound $\max|F_\eta-F_0|$.
For small $a$, waveform deviation and monotonicity limit the strength
to $\delta$ and $\kappa a$, respectively, with $\kappa=4\sqrt3/9$.
The largest allowed strength is set by the smaller bound,
$\eta_{\max}\sim\min(\delta,\kappa a)$.
A narrower feature requires a smaller strength to prevent the total ramp from decreasing locally.
Define the scaled width $x=2\Omega_{\rm mid}a/\epsilon$ and constraint ratio
$r=2\Omega_{\rm mid}\delta/(\kappa\epsilon)$.
Thus $r$ compares the two strength bounds at width $a=\epsilon/(2\Omega_{\rm mid})$.
Maximizing the leading response gives the optimal reflected amplitude
\begin{equation}
 |\beta_*|\sim A_0\epsilon\,x_*\min(r,x_*)e^{-x_*}.
\label{eq:constraintLaw}
\end{equation}
Here $A_0\equiv\pi\Delta_\Omega^2\kappa/(8\Omega_{\rm mid}^3)$ is the
amplitude scale, and the optimal scaled width is $x_* = \min[2,\max(1,r)]$;
stars mark optimal values.
This small-width prediction has three regimes.
For $r<1$, the deviation bound limits $\eta_*\simeq\delta$,
while monotonicity remains inactive; $x_*=1$.
For $1<r<2$, both bounds become active at the optimum, giving $x_*=r$.
For $r>2$, monotonicity limits the strength to $\eta_*\simeq\kappa a_*$,
giving $x_*=2$.
The plateau in Fig.~\ref{fig:design}(d) is in $|\beta_*|/(A_0\epsilon)$:
at fixed $\delta$, the optimal amplitude still decreases as $\epsilon$
and reflected energy as $\epsilon^2$ in the monotonicity-limited regime.
The local adiabatic measure also tends to zero.
For a fixed waveform ($a$ and $\eta$ held constant), increasing $T$ slows
all its structures. Reoptimization with $a\propto\epsilon$ instead keeps
the central feature's time scale, $aT$, of order $1/\omega_0$ as the overall
switching slows.

At $r=0.6,1.5,3$, full mode evolution agrees with Eq.~\eqref{eq:constraintLaw}
within $0.60\%$ for $\epsilon=0.008$ and $0.30\%$ for $0.004$
[Fig.~\ref{fig:design}(d); derivation in the Supplemental Material~\cite{supp}].
The finite-action prediction~\eqref{eq:residue} remains more accurate at the original $\epsilon=0.15$.

\begin{samepage}
\section*{Full-wave validation and bandwidth}

We propagate a Gaussian packet on a one-dimensional Yee lattice~\cite{yee1966}.
Normalizing at its central wave number $k_0$, $\omega_0=\omega_i(k_0)$, gives
$\Omega_i(k)=k/k_0$.
\end{samepage}

The backward branch and separated packet appear in
Figs.~\ref{fig:fullwave}(a) and \ref{fig:fullwave}(b).
Across the resolved bandwidth, the Yee calculation at 80 points per
wavelength agrees with the prediction within $0.31\%$ and with the
60-point grid within $0.025\%$.
The Supplemental Material checks grid refinement, agreement between
oscillator and amplitude equations, and longer simulation windows~\cite{supp}.
With a longer window and matched initialization, the full-wave reference
amplitude at $k_0$ is $8.05\times10^{-10}$, within approximately $0.4\%$ of the analytic result.

The central-mode reflected energy fraction is $(\Omega_f/\Omega_i)|\beta_{\rm opt}|^2=2.30\times10^{-6}$; the field amplitude is of order $10^{-3}$.

Filtering both complete ramps tests the modulation bandwidth needed to retain the central feature. At response time $\tau_b=1/\omega_0$, with inverse $\omega_c=1/\tau_b$, Gaussian smoothing gives $|\beta_\eta|=6.66\times10^{-5}$ and a reference amplitude of $6.54\times10^{-10}$. A causal first-order response at the same $\tau_b$ gives $5.15\times10^{-4}$ and $7.93\times10^{-10}$, respectively~\cite{supp}.

\begin{samepage}
\section*{Discussion}

Weak features can control the residual reflection of a monotonic temporal
interface while leaving its real-time waveform nearly unchanged and its
local adiabatic measure small. The complex action and residue determine
how the feature's time scale and strength affect reflection. The waveform
constraints then select their allowed combinations. This permits a wide
relative variation of the backward output with little change in the
dominant forward amplitude.
\end{samepage}

In adiabatic quantum control, pulse shaping suppresses unwanted transitions
in superconducting-qubit gates by controlling the drive
spectrum~\cite{martinis2014}. Here small waveform changes enhance a weak
reflected component whose scale is not fixed by the local adiabatic measure
alone. Stretching a fixed analytic waveform suppresses reflection
exponentially. Reoptimizing within our family keeps the central feature's
time scale, $aT$, of order $1/\omega_0$, while monotonicity requires its
strength to decrease. At fixed deviation bound $\delta$, the optimal
reflected amplitude then decreases only as a power of the switching time
$T$. The slow limit therefore depends on how the waveform is scaled.

The mode-mixing coefficients also have a quantum interpretation. For the
quantized lossless oscillator, $|\beta_k|^2$ gives the mean occupation
generated from the initial vacuum. The same Bogoliubov mixing describes
parametric amplification and squeezing~\cite{nation2012}. Temporal modulation
has produced photons and two-mode squeezing in superconducting circuits
through the dynamical Casimir effect~\cite{wilson2011}. A question for quantum
temporal interfaces is whether weak switching features can shape
photon-pair production and correlations.

Our quantitative predictions apply to the specified waveform family in
homogeneous, lossless, nondispersive media. The filtering results show why
design must account for the drive response at frequencies relevant to mode
conversion. Including the measured response in the optimization, as in
quantum control~\cite{motzoi2011}, would determine the enhancement achievable
with finite modulation bandwidth. The reflected amplitude and phase could
also probe weak temporal features that are barely visible in the forward
amplitude.

\begin{acknowledgments}
We acknowledge support from the National Key R\&D Program of China
(Grant No.~2023YFA1407001) and the National Natural Science Foundation
of China (Grant No.~12234017).
\end{acknowledgments}

\end{document}


\title{Supplemental Material for ``Exponential Control of Adiabatic Time
Reflection by Complex-Time Singularities''}
\author{Shiyue Deng}
\affiliation{International Center for Quantum Design of Functional Materials (ICQD), Hefei National Research Center for Interdisciplinary Sciences at the Microscale, University of Science and Technology of China, Hefei, Anhui 230026, China}
\affiliation{Hefei National Laboratory, University of Science and Technology of China, Hefei 230088, China}
\author{Qian Niu}
\affiliation{International Center for Quantum Design of Functional Materials (ICQD), Hefei National Research Center for Interdisciplinary Sciences at the Microscale, University of Science and Technology of China, Hefei, Anhui 230026, China}
\affiliation{School of Physics, University of Science and Technology of China, Hefei, Anhui 230026, China}
\date{\today}
\maketitle
\raggedbottom

This supplement derives the complex-time design rule and its constrained
small-width limit. It also gives the reference benchmarks, numerical
checks, drive-response models, and full-wave method.
Wentzel--Kramers--Brillouin theory is abbreviated as WKB below.

\tableofcontents
\clearpage

\section{Maxwell reduction for a spatially uniform temporal medium}

Consider a homogeneous, isotropic, lossless dielectric with
$\bm D=\varepsilon(t)\bm E$ and $\bm B=\mu_0\bm H$. Here $\bm E$ and $\bm H$
are the electric and magnetic fields, $\bm D$ the electric displacement,
and $\bm B$ the magnetic induction. Here $t$ is time, $\varepsilon(t)$
the instantaneous, nondispersive permittivity, and $\mu_0$ the vacuum
permeability. A Fourier component
with wave vector $\bm k$ ($k=|\bm k|$) obeys
\begin{equation}
 \dot{\bm B}_{\bm k}=-i\bm k\times\bm E_{\bm k},\qquad
 \dot{\bm D}_{\bm k}=i\bm k\times\bm H_{\bm k}.
\end{equation}
A dot denotes $d/dt$. Differentiating the second equation and imposing
$\bm k\cdot\bm D_{\bm k}=0$ gives
\begin{equation}
 \ddot{\bm D}_{\bm k}+\frac{k^2}{\mu_0\varepsilon(t)}\bm D_{\bm k}=0.
 \label{eq:MaxwellMode}
\end{equation}
Choose transverse unit vectors $\bm e_{\bm k\lambda}$ and write
$\bm D_{\bm k}=\sum_{\lambda=1,2}D_{\bm k\lambda}\bm e_{\bm k\lambda}$.
Scalar $\varepsilon(t)$ leaves the polarizations uncoupled. We select one and
suppress the polarization index $\lambda$. Let $T$ be the switching time,
$\omega_0$ the reference angular frequency, $\varepsilon_0$ the vacuum
permittivity, and $c=1/\sqrt{\mu_0\varepsilon_0}$ the vacuum light speed.
With dimensionless time $s=t/T$, adiabatic parameter
$\epsilon=(\omega_0T)^{-1}$, and $\varepsilon_r=\varepsilon/\varepsilon_0$,
\begin{equation}
 \epsilon^2D_k''+\Omega_k^2D_k=0,\qquad
 \Omega_k(s)=\frac{ck}{\omega_0\sqrt{\varepsilon_r(s)}}.
 \label{eq:dimensionless}
\end{equation}
A prime denotes $d/ds$. The dimensionless frequency is
$\Omega_k=\omega_k/\omega_0$; subscripts $i,f$ denote initial and final values.
For single-mode theory $\omega_0=\omega_i(k)$, so $\Omega_i=1$;
for a packet centered at wave number $k_0$, $\omega_0=\omega_i(k_0)$ and
$\Omega_i(k)=k/k_0$. We suppress $k$ in single-mode expressions.
Unless stated otherwise, $\Omega_i=1$, $\Omega_f=1.4$, and their squared
difference is $\Delta_\Omega^2\equiv\Omega_f^2-\Omega_i^2=0.96$.

At constant dimensionless frequency $\Omega$, define normalized
positive- and negative-frequency modes
\begin{equation}
 u_\Omega^{(\pm)}=\frac{e^{\mp i\Omega s/\epsilon}}{\sqrt{2\Omega}}.
\end{equation}
The Wronskian $W[f,g]=fg'-f'g$ is constant for any two solutions of
Eq.~\eqref{eq:dimensionless}, and
$\epsilon W[u_\Omega^{(+)},u_\Omega^{(-)}]=i$. Thus
$D_k=\alpha_k u_{\Omega_f}^{(+)}+\beta_k u_{\Omega_f}^{(-)}$, where
$\alpha_k$ and $\beta_k$ are the transmitted and reflected amplitudes.
For unit positive-frequency input they satisfy
\begin{equation}
 |\alpha_k|^2-|\beta_k|^2=1.
 \label{eq:wronskianNorm}
\end{equation}
At conserved $k$, the negative-frequency component has the opposite group
velocity and is the time-reflected wave.

\section{Exact reference interface}

The solvable background is
\begin{equation}
 F_0(s)=\frac{1+\tanh s}{2},\qquad
 \Omega^2=\Omega_i^2+\Delta_\Omega^2F_0(s).
 \label{eq:tanhRamp}
\end{equation}
At fixed $k$, $\Omega^2\propto1/\varepsilon_r$, so this choice interpolates
the inverse relative permittivity as
$1/\varepsilon_r(s)=[1-F_0(s)]/\varepsilon_{r,i}+F_0(s)/\varepsilon_{r,f}$,
where $\varepsilon_{r,i}$ and $\varepsilon_{r,f}$ are its initial and final
relative permittivities. The permittivity itself is not a tanh ramp.
Use the auxiliary coordinate $\zeta(s)\equiv F_0(s)$, so
$d\zeta/ds=2\zeta(1-\zeta)$. Factoring the endpoint phases
reduces Eq.~\eqref{eq:dimensionless} to the Gauss hypergeometric equation.
Define the endpoint exponents
$\gamma_i=-i\Omega_i/(2\epsilon)$ and $\gamma_f=i\Omega_f/(2\epsilon)$.
The unit positive-frequency input solution is
\begin{equation}
 D_{\rm in}(s)=\frac{\zeta^{\gamma_i}(1-\zeta)^{\gamma_f}}{\sqrt{2\Omega_i}}
 \,{}_2F_1\!\left(\gamma_i+\gamma_f,\gamma_i+\gamma_f+1;
 1+2\gamma_i;\zeta\right),\qquad \zeta=\zeta(s),
\end{equation}
where ${}_2F_1$ is the Gauss hypergeometric function. Its continuation to
$\zeta\to1$ contains the negative-frequency factor
$(1-\zeta)^{-\gamma_f}$, giving
\begin{equation}
 \beta_0=\sqrt{\frac{\Omega_f}{\Omega_i}}
 \frac{\Gamma(1+2\gamma_i)\Gamma(2\gamma_f)}
 {\Gamma(\gamma_i+\gamma_f)\Gamma(1+\gamma_i+\gamma_f)}.
\end{equation}
Here $\Gamma$ is the Euler gamma function. Taking the modulus with
$|\Gamma(ix)|^2=\pi/[x\sinh(\pi x)]$ for real $x\ne0$ yields
\begin{equation}
 |\beta_0|^2=
 \frac{\sinh^2[\pi(\Omega_f-\Omega_i)/(2\epsilon)]}
 {\sinh(\pi\Omega_i/\epsilon)\sinh(\pi\Omega_f/\epsilon)}.
 \label{eq:exactBeta}
\end{equation}
This is exact for arbitrary $\epsilon$. For unequal endpoints its limits are
\begin{align}
 |\beta_0|&\sim e^{-\pi\min(\Omega_i,\Omega_f)/\epsilon}
 &&(\epsilon\to0),\label{eq:tanhAsymptotic}\\
 |\beta_0|^2&\to\frac{(\Omega_f-\Omega_i)^2}{4\Omega_i\Omega_f}
 &&(\epsilon\to\infty).
\end{align}

The complex turning points, indexed by the integer $n$, are
\begin{equation}
 s_c^{(n)}=\ln\frac{\Omega_i}{\Omega_f}
 +i\pi\left(n+\frac12\right).
\end{equation}
For our endpoints, on the sheet connected to real time, the nearest upper turning point
$s_c=s_c^{(0)}$ gives the action
\begin{equation}
 S_0=2\operatorname{Im}\int_{\operatorname{Re}s_c}^{s_c}\Omega(s)\,ds=\pi.
 \label{eq:tanhAction}
\end{equation}
The hypergeometric solution gives Eq.~\eqref{eq:exactBeta} at any switching
rate; complex-time analysis gives its adiabatic exponent. The exact result
provides a benchmark for the exponentially small WKB remainder in
Fig.~\ref{fig:optimalTruncation} below.

\section{From adiabatic amplitudes to complex time}

For an integration interval $[-L,L]$, with $L>0$ large enough to reach
the asymptotic tails, write the exact solution in an instantaneous frequency basis,
\begin{equation}
 D=\frac{\alpha e^{-i\theta/\epsilon}+\beta e^{i\theta/\epsilon}}
 {\sqrt{2\Omega}},\qquad
 \theta(s)=\int_{-L}^{s}\Omega(x)\,dx,
 \label{eq:instantaneousBasis}
\end{equation}
where $\theta$ is the accumulated dimensionless phase. Define the amplitudes
also through the field derivative,
\begin{equation}
 \epsilon D'=-i\sqrt{\frac{\Omega}{2}}
 [\alpha e^{-i\theta/\epsilon}-\beta e^{i\theta/\epsilon}].
\end{equation}
These equations define $\alpha,\beta$ uniquely. Differentiating them and
using $\epsilon^2D''=-\Omega^2D$ gives
\begin{align*}
 \alpha'e^{-i\theta/\epsilon}+\beta'e^{i\theta/\epsilon}
 &=\frac{\Omega'}{2\Omega}
 (\alpha e^{-i\theta/\epsilon}+\beta e^{i\theta/\epsilon}),\\
 \alpha'e^{-i\theta/\epsilon}-\beta'e^{i\theta/\epsilon}
 &=-\frac{\Omega'}{2\Omega}
 (\alpha e^{-i\theta/\epsilon}-\beta e^{i\theta/\epsilon}).
\end{align*}
Their sum and difference give the exact coupled equations,
\begin{align}
 \alpha'&=\frac{\Omega'}{2\Omega}e^{2i\theta/\epsilon}\beta,
 \nonumber\\
 \beta'&=\frac{\Omega'}{2\Omega}e^{-2i\theta/\epsilon}\alpha.
 \label{eq:BogoliubovODE}
\end{align}
The coupling is $\Omega'/(2\Omega)=(\Omega^2)'/(4\Omega^2)$. The relative
phase between the branches accumulates at $2\omega(t)$, which accounts
for the factor of 2 in the complex action.
For $\beta(-L)=0$,
\begin{equation}
 \beta(L)=\int_{-L}^{L}\frac{\Omega'}{2\Omega}
 e^{-2i\theta/\epsilon}\alpha(s)\,ds.
 \label{eq:betaIntegralExact}
\end{equation}
The first iteration sets $\alpha\simeq1$, neglecting feedback and the
associated phase correction. To see the cancellation, write
$g(s)=\Omega'(s)/(2\Omega(s))$ and
$\phi(s)=e^{-2i\theta(s)/\epsilon}$. Since
$\phi'=-2i\Omega\phi/\epsilon$, integration by parts gives
\begin{equation}
 \int_{-L}^{L}g\phi\,ds
 =-\frac{\epsilon}{2i}
 \left[\frac{g}{\Omega}\phi\right]_{-L}^{L}
 +\frac{\epsilon}{2i}
 \int_{-L}^{L}\left(\frac{g}{\Omega}\right)'\phi\,ds.
 \label{eq:integrationByParts}
\end{equation}
For these ramps, $\Omega>0$ and the endpoint derivatives decay, so the
boundary term vanishes as $L\to\infty$. Repeated integration by parts gives
further powers of $\epsilon$ because there is no real stationary point.
This first iteration illustrates the cancellation; finite-order WKB also
leaves no algebraic outgoing backward amplitude. Analytic continuation
determines the exponential scale, which the exact tanh solution checks
independently.
Derivative jumps can instead leave algebraic boundary contributions
(M. V. Berry,
\href{https://doi.org/10.1088/0305-4470/15/12/021}{J. Phys. A \textbf{15}, 3693 (1982)}).
A frequency jump gives the finite temporal-step reflection.

Let $\Omega_0(s)$ denote the frequency of the $F_0$ reference, and define
its phase
\begin{equation}
 \chi_0(s)=2i\int_0^s\Omega_0(u)\,du.
 \label{eq:complexPhase}
\end{equation}
At finite time the boundary term can be $O(\epsilon)$. For the tanh ramp
at $\epsilon=0.15$, direct integration gives the instantaneous-basis
$|\beta(0)|=5.03\times10^{-3}$ but final $|\beta|=8.02\times10^{-10}$:
later contributions cancel most of the mixing during switching.

Continue $\Omega_0$ from its positive real-axis value along a specified
contour. Shifting the real reference point leaves the action's imaginary
part unchanged, provided the frequency branch and contour winding are kept
fixed.
At a simple zero, $\Omega_0^2\propto s-s_c$: its square root is branched
and WKB fails although the oscillator equation is regular. At a simple pole,
$\Omega_0^2\propto(s-s_c)^{-1}$ makes the equation singular. The coupling
has residues $+1/4$ and $-1/4$, respectively. The oscillatory phase selects
the lower half-plane, with contributing features determined by the Stokes
geometry. Turning points require a local WKB connection; the perturbation
pole below admits a contour-residue calculation. Stokes switching concerns
the asymptotic representation, while real-time evolution remains smooth.

\section{Exact first variation and the complex-pole rule}

For perturbation strength $\eta\geq0$ and dimensionless feature width $a>0$, use
\begin{equation}
 F_\eta=F_0+\eta h_a,\qquad
 h_a=\frac{s^2}{s^2+a^2}\operatorname{sech}^2s,\qquad
 \Omega^2=\Omega_0^2+\eta\Delta_\Omega^2h_a,
 \label{eq:designedFamily}
\end{equation}
Here $h_a$ is the perturbation profile and $\operatorname{sech}s=1/\cosh s$.
The factor $s/a=t/(aT)$ sets the central feature's time scale $aT$,
whereas the envelope varies on scale $T$. Width $a$ and strength $\eta$
are the design parameters.
Write $V(s)=\Delta_\Omega^2h_a(s)$ for the frequency-squared perturbation per
unit $\eta$, and $\Delta\beta=\beta_\eta-\beta_0$ for the reflection change.
Let $u_{\rm in}^{(0)}$ be the exact
reference solution with unit positive-frequency input, and let
$u_{{\rm out},+}^{(0)}$ be the exact reference solution with unit pure
positive-frequency output. The phases of these two solutions are fixed by
\begin{equation}
 u_{\rm in}^{(0)}\sim\frac{e^{-i\Omega_i s/\epsilon}}{\sqrt{2\Omega_i}}
 \quad(s\to-\infty),\qquad
 u_{{\rm out},+}^{(0)}\sim\frac{e^{-i\Omega_f s/\epsilon}}{\sqrt{2\Omega_f}}
 \quad(s\to+\infty).
 \label{eq:referenceModes}
\end{equation}
We compare scattering coefficients in these common asymptotic phases.
For the field variation and reflection response at fixed $\epsilon$, define
$v=\left.\partial_\eta u_{\rm in}^{(\eta)}\right|_{\eta=0}$ and
$B=\left.\partial_\eta\beta_\eta\right|_{\eta=0}$.
Differentiating the oscillator gives
\begin{equation}
 \epsilon^2v''+\Omega_0^2v=-Vu_{\rm in}^{(0)},\qquad
 v\to0\quad(s\to-\infty).
 \label{eq:variationalOscillator}
\end{equation}
The Wronskian defined above satisfies
\begin{equation}
 \frac{d}{ds}W[u_{{\rm out},+}^{(0)},v]
 =-\frac{V}{\epsilon^2}u_{{\rm out},+}^{(0)}u_{\rm in}^{(0)}.
\end{equation}
At the output the Wronskian is $iB/\epsilon$. Integrating gives an exact
identity for the derivative,
\begin{equation}
 B=\frac{i}{\epsilon}\int_{-\infty}^{\infty}
 V(s)u_{{\rm out},+}^{(0)}(s)u_{\rm in}^{(0)}(s)\,ds,
 \qquad \beta_\eta-\beta_0=\eta B+O(\eta^2).
 \label{eq:exactVariation}
\end{equation}
The integral gives $B$ exactly; the finite change $\eta B$ is a first-order
approximation.
At fixed shape, $\eta/\epsilon\ll1$ controls the accumulated phase correction.
The shrinking-feature limit below treats $\eta=O(\epsilon)$ separately.

For the design rule, approximate the two background solutions by leading
WKB. Define $\theta_0(s)=\int_0^s\Omega_0(u)\,du$ and the finite
asymptotic phase constants
\begin{equation}
 \varphi_i=\lim_{s\to-\infty}[\theta_0(s)-\Omega_i s],\qquad
 \varphi_f=\lim_{s\to+\infty}[\theta_0(s)-\Omega_f s].
\end{equation}
Then their product is approximately
$e^{i(\varphi_i+\varphi_f)/\epsilon}e^{-2i\theta_0(s)/\epsilon}/[2\Omega_0(s)]$,
and Eq.~\eqref{eq:exactVariation} becomes
\begin{equation}
 B_{\rm WKB}=\frac{i}{2\epsilon}
 e^{i(\varphi_i+\varphi_f)/\epsilon}
 \int_{-\infty}^{\infty}\frac{V(s)}{\Omega_0(s)}
 e^{-\chi_0(s)/\epsilon}\,ds.
 \label{eq:betaPerturbationIntegral}
\end{equation}
At the lower pole $s_p=-ia$, $V$ has residue $r_a$:
\begin{equation}
 V(s)=\frac{r_a}{s-s_p}+O(1),\qquad
 r_a=-\frac{i\Delta_\Omega^2a}{2}\sec^2a.
\end{equation}
For $0<a<\pi/2$, shift the contour to $s=x-id$, with real $x$ and $a<d<\pi/2$.
The reference solutions are analytic in this strip and $V$ decays at its ends.
The clockwise contour around the lower pole gives
\begin{equation}
 B=\frac{2\pi r_a}{\epsilon}
 u_{{\rm out},+}^{(0)}(s_p)u_{\rm in}^{(0)}(s_p)+R_d,
 \quad R_d=\frac{i}{\epsilon}\int_{-\infty}^{\infty}
 [Vu_{{\rm out},+}^{(0)}u_{\rm in}^{(0)}](x-id)\,dx.
 \label{eq:exactContour}
\end{equation}
When the shifted-contour remainder satisfies $|R_d|\ll|B|$, leading WKB in the pole
kernel gives the complex-time (CT) prediction
\begin{equation}
 B_{\rm CT}=\frac{\pi r_a}{\epsilon\Omega_0(s_p)}
 \exp\left[\frac{i(\varphi_i+\varphi_f)-\chi_0(s_p)}{\epsilon}\right].
 \label{eq:complexPrediction}
\end{equation}
Its magnitude gives
\begin{equation}
 |\beta_\eta-\beta_0|\sim\eta\frac{C(a)}{\epsilon}e^{-S(a)/\epsilon},
 \label{eq:residueFormula}
\end{equation}
where
\begin{align}
 S(a)&=2\operatorname{Im}\int_0^{ia}\Omega_0(s)\,ds,
 \label{eq:poleAction}\\
 C(a)&=\frac{\pi|r_a|}{|\Omega_0(-ia)|}
 =\frac{\pi\Delta_\Omega^2a\sec^2a}{2|\Omega_0(-ia)|}.
 \label{eq:polePrefactor}
\end{align}
On the frequency branch connected to real time,
$\Omega_0(s^*)=\Omega_0(s)^*$, so $\operatorname{Re}\chi_0(-ia)=S(a)$:
the lower pole and its upper conjugate give the same positive action.
The phase determines $S$, the residue determines $C$, and the exact
variation supplies $1/\epsilon$. When several complex contributions have
comparable magnitudes, their amplitudes must be summed coherently.

The complete $\Omega_\eta^2$ also has a nearby zero $s_z$. Near $s_p$,
$\Omega_\eta^2\simeq\Omega_0^2(s_p)+\eta r_a/(s-s_p)$, so
\begin{equation}
 s_z-s_p\simeq-\frac{\eta r_a}{\Omega_0^2(s_p)}.
 \label{eq:poleZeroPair}
\end{equation}
The zero approaches the pole as $\eta\to0$, removing the added structure.
The exact coupling has simple poles at both positions:
\begin{equation*}
 \frac{(\Omega_\eta^2)'}{4\Omega_\eta^2}
 =\frac14\left(\frac{1}{s-s_z}-\frac{1}{s-s_p}\right)
 +\text{terms regular at both points}.
\end{equation*}
Expanding for $|s_z-s_p|\ll|s-s_p|$ gives a second-order pole in the
coupling; the full $\Omega_\eta^2$ retains a simple zero and simple pole.
Equation~\eqref{eq:exactVariation} includes their combined first-order
response, so no additional turning-point term is added. At the selected
design their separation is $1.56\times10^{-4}\ll a$.
The action in Eq.~\eqref{eq:poleAction} is that of the reference kernel.

For $a=0.070$ and $\eta=0.00687974$, $S=0.170321$ and $C=0.0871830$.
At $\epsilon=0.15$, $\eta|B_{\rm CT}|=1.284645\times10^{-3}$,
$\eta|B|=1.283923\times10^{-3}$, and
$|\beta_\eta|=1.280549\times10^{-3}$. The reference is negligible here.
The WKB-pole amplitude differs from the exact linear response by $0.0563\%$.
The complex relative difference is $1.04\%$, mainly from the phase error.
Using exact reference solutions in the pole kernel gives a complex relative
difference of $1.4\times10^{-7}$ from $B$.
These comparisons distinguish the background WKB error, the contour
remainder, and the finite-$\eta$ correction. We also test the same selected
waveform at seven values of $\epsilon$ between $0.10$ and $0.30$.
The predicted $|\Delta\beta|$ differs from direct integration by at most
$0.40\%$, without refitting any parameter.

\section{Real-axis constraints and optimization}

The family preserves both endpoints, the midpoint, and the central slope.
Monotonicity retains a single switch without material-parameter reversals.
Write $H(s)=\operatorname{sech}^2s$ and define the derivative factor $b_a(s)$ by
\begin{equation}
 \begin{aligned}
 F_\eta'(s)&=H(s)[\tfrac12+\eta b_a(s)],\\
 b_a(s)&=\frac{2sa^2}{(s^2+a^2)^2}
 -\frac{2s^2\tanh s}{s^2+a^2}.
 \end{aligned}
 \label{eq:continuousConstraint}
\end{equation}
For real $s$, $|2sa^2/(s^2+a^2)^2|\leq9/(8\sqrt3a)$ and
$|\tanh s|\leq1$, giving the conservative bound
\begin{equation}
 F_\eta'\geq\operatorname{sech}^2s
 \left[\frac12-\eta\left(2+\frac{9}{8\sqrt3a}\right)\right].
 \label{eq:monotonicBound}
\end{equation}
At the selected design the bracket is $0.4224>0$, proving global
monotonicity. Let $\delta$ bound $\max|F_\eta-F_0|$. The exact allowed strength is
\begin{equation}
 \eta_{\max}(a,\delta)=\min\left[
 \frac{\delta}{\max_{s\in\mathbb R}h_a(s)},
 -\frac{1}{2\inf_{s\in\mathbb R}b_a(s)}\right].
 \label{eq:exactConstraint}
\end{equation}
The infimum includes the tail limit $b_a\to-2$ as $s\to+\infty$ and is
negative. The second entry defines the monotonicity-limited strength
$\eta_{\rm mono}$. We determine the continuous extrema and check the
derivative on a fine grid. At fixed width the leading reflection increases
with $\eta$, so optimization uses the smaller strength bound.
For $\delta=0.006$, maximizing Eq.~\eqref{eq:residueFormula}
over $0.035\leq a\leq0.8$ gives
$a\simeq0.07055$; we use $a=0.070$, whose predicted amplitude differs by
only $0.0031\%$. This is a near optimum within the specified family.
The local adiabatic measure
$\mathcal A(s)=\epsilon|\Omega'|/\Omega^2=|\dot\omega|/\omega^2$
measures the fractional frequency change over a time interval $1/\omega$.
Its peak is $0.0220720$, versus $0.0212798$ for the reference ramp.
The narrow feature remains weak enough to keep this measure small.

The full width scan is retained in the
\href{../../outputs/complex_time_inverse_design_exact_points.csv}{design data}.

\section{Width dependence and an alternative envelope}

For a fixed analytic waveform, increasing $T$ slows every feature and
exponentially suppresses final reflection. At our selected point,
$a/\epsilon=0.467$ and $S/\epsilon=1.14$; the small output depends on
both the exponential factor, about $0.32$, and the weak perturbation strength.

For $a\ll1$, define the midpoint frequency $\Omega_{\rm mid}=\sqrt{(\Omega_i^2+\Omega_f^2)/2}$.
The added pole at $ia$ is a distance $a$ from the switching center $s=0$
in the complex-time plane. The background frequency varies little along
the short contour between these points.
Expanding the action and prefactor gives
\begin{align*}
 S(a)&=2\Omega_{\rm mid}a+O(a^3),\\
 C(a)&=\frac{\pi\Delta_\Omega^2a}{2\Omega_{\rm mid}}[1+O(a^2)].
\end{align*}
Here the term linear in $s$ in the frequency expansion contributes only
to the real part of the contour integral. Substitution into
Eq.~\eqref{eq:residueFormula} gives the leading small-width rule,
\begin{equation}
 \eta|B_{\rm CT}|\simeq
 \frac{\eta\pi\Delta_\Omega^2}{2\Omega_{\rm mid}}
 \frac{a}{\epsilon}e^{-2\Omega_{\rm mid}a/\epsilon}.
 \label{eq:smallAScaling}
\end{equation}
The action and residue both decrease with width. Their competition gives
$a\simeq\epsilon/(2\Omega_{\rm mid})$ at fixed strength.
The constraints change the strength with width and shift this optimum;
their shrinking-feature limit is derived in the next section.

\begin{figure}[H]
 \centering
 \includegraphics[width=0.82\textwidth]{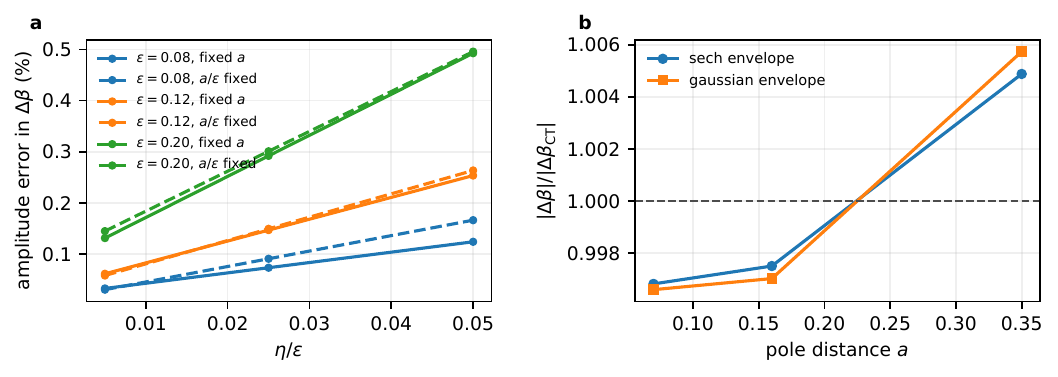}
 \caption{\label{fig:weakResponse}Accuracy and envelope dependence of the design rule.
 (a) Errors for fixed pole distance (solid) and fixed $a/\epsilon$
 (dashed), with different $\epsilon$ and $\eta/\epsilon$.
 (b) Exact reflection change divided by its predicted amplitude for two
 envelopes with the same background and pole positions. Changing the envelope
 changes the residue but leaves the background action unchanged. Neither
 envelope requires fitting.}
\end{figure}

In this test, $\eta$ is varied independently of the design optimization.
We test the approximation at $\epsilon=0.08,0.12,0.20$ and
$\eta/\epsilon=0.005,0.025,0.05$, both at fixed $a=0.070$ and at fixed
$a/\epsilon=0.467$. These 18 finite-parameter checks give a maximum
action--residue amplitude difference of $0.50\%$ from $|\Delta\beta|$.
The exact first variation differs from the finite-$\eta$ complex
$\Delta\beta$ by at most $0.54\%$. Figure~\ref{fig:weakResponse}(a)
shows the amplitude errors. These finite-parameter checks do not establish
a uniform fixed-$\eta$ asymptotic theorem.

To test dependence on the envelope, replace $h_a$ by
\begin{equation}
 \widetilde h_a(s)=\frac{s^2}{s^2+a^2}e^{-2s^2},\qquad
 F_\eta=F_0+\eta\widetilde h_a.
 \label{eq:secondEnvelope}
\end{equation}
The endpoints, midpoint, and central slope are preserved. The background
action $S(a)$ is unchanged, while the residue becomes
$\widetilde r_a=-i\Delta_\Omega^2a e^{2a^2}/2$ and hence
$\widetilde C=\pi|\widetilde r_a|/|\Omega_0(-ia)|$.
For each $a$ choose $\eta\max\widetilde h_a=0.006$.
Global monotonicity follows from a conservative bound. Since
$e^{-2s^2}\cosh^2s\leq e^{1/2}$ and
$4|s|e^{-2s^2}\cosh^2s<4.42$,
\begin{equation}
 \frac{F_\eta'}{F_0'}\geq
 1-2\eta\left[e^{1/2}\frac{9}{8\sqrt3a}+4.42\right]>0
 \label{eq:gaussianMonotonicity}
\end{equation}
at all three tested points. Their peak local adiabatic measures lie between
$0.0205$ and $0.0223$. Figure~\ref{fig:weakResponse}(b) tests
$a=0.070,0.160,0.350$ for both envelopes without fitting; amplitude
 errors are at most $0.58\%$. The values for each design are retained in the
\href{../../outputs/weak_response_envelopes.csv}{accompanying envelope data}.

\section{Constraint competition in the shrinking-feature limit}

We take $a,\eta=O(\epsilon)$ in the specified family, under
$\max|F_\eta-F_0|\leq\delta$. At the optima below, $a/\epsilon$ tends to a
positive constant, so the central duration $aT$ remains comparable to
$1/\omega_0$ while the overall switching slows. This differs from stretching
a fixed waveform.

\subsection{Continuous monotonicity and deviation bounds}

We expand the exact constraints in Eq.~\eqref{eq:exactConstraint}.
In the local coordinate $\xi=s/a$,
$b_a(a\xi)=a^{-1}2\xi/(1+\xi^2)^2+O(a)$ for bounded $\xi$.
The first term has its minimum at $\xi=-1/\sqrt3$, with value
$-9/(8\sqrt3a)$. Contributions away from this neighborhood remain
smaller than this diverging minimum as $a\to0$. Consequently
\begin{equation}
 \eta_{\rm mono}=\kappa a[1+O(a^2)],\qquad
 \kappa=\frac{4\sqrt3}{9}.
 \label{eq:monoScaling}
\end{equation}
For the deviation constraint, the maximum occurs at
$s^2=a[1+O(a)]$, as follows by balancing
$h_a=1-s^2-a^2/s^2+\cdots$. Hence
$\max h_a=1-2a+O(a^2)$ and
the deviation-limited strength is $\delta[1+2a+O(a^2)]$.
To leading order, $\eta_{\max}=\min(\delta,\kappa a)$: the allowed strength
must satisfy both bounds. A narrower feature permits a smaller strength
because its negative derivative would otherwise overwhelm the background slope.

\subsection{Retaining the slow background avoids a secular phase expansion}

At the monotonicity limit, $\eta/\epsilon$ can approach a nonzero constant.
The fixed-$\epsilon$ Taylor expansion of the complete scattering phase is
then unsuitable for establishing a limiting amplitude. Instead use the
exact decomposition
\begin{equation}
 \begin{aligned}
 \Omega_\eta^2(s)&=\Omega_b^2(s)+U_a(s),\\
 \Omega_b^2(s)&=\Omega_0^2(s)+\eta\Delta_\Omega^2H(s),\\
 U_a(s)&=-\eta\Delta_\Omega^2
              \frac{a^2H(s)}{s^2+a^2}.
 \end{aligned}
 \label{eq:resummedBackground}
\end{equation}
$U_a$ is the localized frequency-squared notch. The retained background
frequency $\Omega_b$ is positive and smooth, has fixed-scale analytic
structure, and preserves the endpoints. Its reflection is exponentially
small, uniformly for sufficiently small $\eta$. We retain the background
phase exactly and treat only $U_a$ by weak scattering.
Its norm satisfies
\begin{equation}
 \frac{1}{\epsilon}\int_{\mathbb R}|U_a(s)|\,ds
 \leq\frac{\pi\Delta_\Omega^2\eta a}{\epsilon}=O(\epsilon).
 \label{eq:notchNorm}
\end{equation}
Choose a unit positive-frequency background output mode $u_b^+$ and
its complex conjugate $u_b^-=(u_b^+)^*$, with $W_b=W[u_b^+,u_b^-]=i/\epsilon$.
For dimensionless times $s\geq u$ define the retarded Green kernel
\begin{equation}
 G_b(s,u)=\frac{u_b^+(u)u_b^-(s)-u_b^-(u)u_b^+(s)}{W_b}.
 \label{eq:backgroundGreen}
\end{equation}
It has $G_b(u,u)=0$ and
$\left.\partial G_b(s,u)/\partial s\right|_{s=u}=1$.
If $u_{b,\rm in}$ is the exact background input solution, the complete
input mode satisfies
\begin{equation}
 D(s)=u_{b,\rm in}(s)-\frac{1}{\epsilon^2}
             \int_{-\infty}^s G_b(s,u)U_a(u)D(u)\,du.
 \label{eq:notchVolterra}
\end{equation}
Uniform real-axis WKB estimates for the positive, fixed-scale background
bound $u_b^\pm$ and $\epsilon\,du_b^\pm/ds$ by constants independent of
small $\epsilon$. Thus $|G_b|\leq M\epsilon$ and
$|\partial G_b(s,u)/\partial s|\leq M$, with $M$ a positive bound
independent of $a=O(\epsilon)$.
In the norm $|D|+\epsilon|D'|$, the integral operator has size at most
$(M/\epsilon)\int|U_a|$. Its Neumann series and the endpoint projection
therefore bound the second and higher scattering terms by
$O(\epsilon^2)$, for $a/\epsilon$ and $\eta/\epsilon$ bounded.
This estimate controls the localized scattering and retains the complete
global background phase.

Let the width and strength ratios approach constants,
$a/\epsilon\to \rho>0$ and $\eta/\epsilon\to m>0$.
In the local coordinate $y=s/\epsilon$, the background mode product, after removal of
its real endpoint phase, tends to
$e^{-2i\Omega_{\rm mid}y}/(2\Omega_{\rm mid})$.
Using $H(\epsilon y)\to1$ in the first-scattering integral and dominated
convergence with $\rho^2/(y^2+\rho^2)$ gives
\begin{equation}
 \frac{|\beta_\eta|}{\epsilon}\longrightarrow
 \frac{\pi\Delta_\Omega^2}{2\Omega_{\rm mid}}
       m\rho\,e^{-2\Omega_{\rm mid}\rho}.
 \label{eq:resummedLimit}
\end{equation}
Here we used
$\int_{\mathbb R}\rho^2e^{-ipy}/(y^2+\rho^2)\,dy=\pi \rho e^{-\rho|p|}$
for real Fourier variable $p$.
The background reflection is negligible on this scale.
The local WKB limit of the kernel and the integrable notch bound justify
this leading limit. They do not give a uniform relative-error rate for
arbitrary $\rho,m$. The real global phase may vary by order one even when
the magnitude converges. Thus the amplitude result does not require a
small-$\eta/\epsilon$ expansion.

\subsection{Three regimes and their numerical checks}

Define the scaled width $x=2\Omega_{\rm mid}a/\epsilon$,
constraint ratio
$r=2\Omega_{\rm mid}\delta/(\kappa\epsilon)$, and amplitude scale
$A_0\equiv\pi\Delta_\Omega^2\kappa/(8\Omega_{\rm mid}^3)$.
The ratio $r$ compares the waveform-deviation bound $\delta$ with the
monotonicity bound at width $a=\epsilon/(2\Omega_{\rm mid})$.
At any width $x$, the ratio of these two strength bounds is $r/x$.
Equations~\eqref{eq:monoScaling} and \eqref{eq:resummedLimit} reduce
the constrained leading response to
\begin{equation}
 \frac{|\beta|}{A_0\epsilon}
   \sim x\min(r,x)e^{-x}.
 \label{eq:constraintFunction}
\end{equation}
For $x<r$ the strength is limited by monotonicity and the function is
$x^2e^{-x}$; for $x>r$ the deviation constraint gives $rxe^{-x}$.
Their derivatives vanish at $x=2$ and $x=1$, respectively. Comparing the
two intervals gives the unique leading maximum (stars denote optimal values)
\begin{equation}
 \begin{array}{c|c|c}
 \text{range}&x_*&|\beta_*|/(A_0\epsilon)\\\hline
 r\leq1&1&r/e\\
 1<r<2&r&r^2e^{-r}\\
 r\geq2&2&4e^{-2}
 \end{array}
 \label{eq:threeRegimes}
\end{equation}
The optimum is deviation-limited for $r<1$, reaches both bounds for
$1<r<2$, and is monotonicity-limited for $r>2$. The boundaries $r=1,2$
belong to the leading small-width problem; finite-width corrections shift
the crossovers. This optimum concerns the specified family, not all temporal profiles.
For fixed positive $r$, Eq.~\eqref{eq:resummedLimit} applies as
$\epsilon\to0$ with $\delta=r\kappa\epsilon/(2\Omega_{\rm mid})$.
For fixed positive $\delta$, the monotonicity branch eventually becomes
active, also with $a,\eta=O(\epsilon)$, giving
$|\beta_*|\sim4A_0e^{-2}\epsilon$.
The plateau $4e^{-2}$ is in the scaled amplitude: the unscaled optimum
still vanishes as $\epsilon$. Since $F_\eta'$ stays bounded,
the local adiabatic measure is $O(\epsilon)$. The central duration remains
of order $1/\omega_0$, but its strength decreases as the overall switching
slows, giving a reflected energy fraction of order $\epsilon^2$.

\begin{figure}[!htbp]
 \centering
 \includegraphics[width=0.90\textwidth]{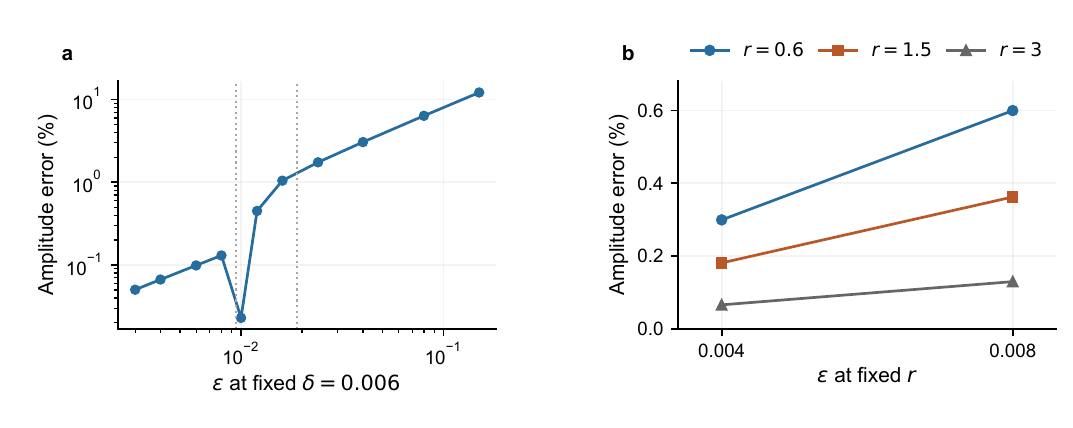}
 \caption{\label{fig:crossover}Convergence to the constrained small-width law.
 Relative amplitude error is
 $\left|\,|\beta|_{\rm leading}/|\beta|_{\rm direct}-1\right|$;
 $|\beta|_{\rm leading}$ is the piecewise asymptotic prediction and
 $|\beta|_{\rm direct}$ comes from full mode integration at the finite-action
 selected design. (a) All eleven switching rates at fixed $\delta=0.006$;
 the two vertical lines mark the leading regime boundaries.
 (b) Fixed $r=0.6,1.5,3$, one in each regime, with
 $\delta=r\kappa\epsilon/(2\Omega_{\rm mid})$. Halving $\epsilon$ reduces
 the error approximately by half. Lines connect computed points.}
\end{figure}

For $\delta=0.006$, the leading boundaries are
$\epsilon=\Omega_{\rm mid}\delta/\kappa\simeq0.00948$ and
$2\Omega_{\rm mid}\delta/\kappa\simeq0.01896$.
Finite-width corrections shift these values, and the finite-action formula
remains preferable to the small-scale limit at $\epsilon=0.15$.
We optimize that formula with the exact continuous constraints, then
independently propagate the selected mode. Eleven values of $\epsilon$
from $0.003$ to $0.15$ pass monotonicity and deviation checks.
At each selected point, exact amplitudes at $0.8a_*$ and $1.2a_*$ are
smaller; these are local consistency checks, not an exact global numerical
optimization. Figure~\ref{fig:crossover}(a) shows the amplitude error
against the leading law at every tested switching rate.

To test convergence to the limit, hold $r=0.6,1.5,3$ fixed and halve
$\epsilon$ from $0.008$ to $0.004$. The relative amplitude error decreases
approximately by a factor of two in each regime
[Fig.~\ref{fig:crossover}(b)]. This tests convergence to each of the three
branches beyond agreement at a single design.
Using a longer integration interval and tighter tolerances at
$\epsilon=0.004,0.012$ changes the amplitudes by less than
$7.3\times10^{-9}$ relatively. An independent original second-order
oscillator integration at $\epsilon=0.012$ agrees with the mode-equation
amplitude to $7.4\times10^{-9}$ relatively.
The code and retained data specify the
finite-scale selection and tests.

\section{Divergent adiabatic series and least-term truncation}

For the reference ramp, the scaled logarithmic derivative $q=\epsilon D'/D$
obeys $q^2+\epsilon q'+\Omega^2=0$. Expanding
$q\sim\sum_{n=0}^\infty\epsilon^nq_n$, with $q_0=-i\Omega$, gives
\begin{equation}
 q_n=-\frac{q_{n-1}'+\sum_{j=1}^{n-1}q_jq_{n-j}}{2q_0},\qquad n\geq1.
\end{equation}
We compute through $n=70$ at $s_{\rm obs}=2$ with 100-digit precision.
The singulant, the complex action governing late terms, is
\begin{equation}
 \chi(s_{\rm obs})=2i\int_{s_c}^{s_{\rm obs}}\Omega(s)\,ds.
\end{equation}
Here $\chi=\pi+6.7545554i$: $\operatorname{Re}\chi=S_0$ controls reflection,
while $|\chi|=7.4494042$ gives the least-term scale $N_\ast\sim|\chi|/\epsilon$.

\begin{figure}[!htbp]
 \centering
 \includegraphics[width=0.90\textwidth]{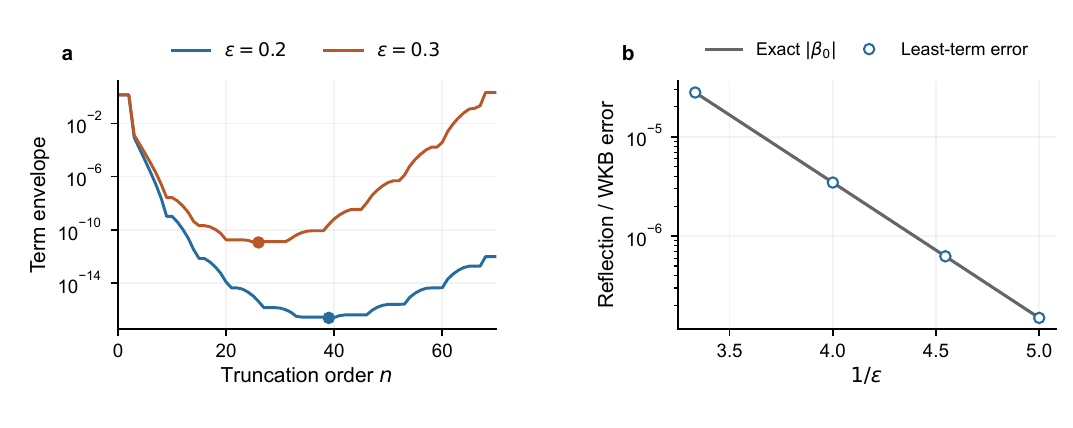}
 \caption{\label{fig:optimalTruncation}Exponentially small reflection beyond
 least-term WKB at $s_{\rm obs}=2$. (a) Five-point upper envelopes of
 $|\epsilon^nq_n|$ at two representative switching rates. Circles mark
 $N_\ast$ selected from the terms, before comparison with the exact solution.
 (b) The normalized error $E_{N_\ast}$ follows the exact tanh reflection
 $|\beta_0|$; their relative difference is below $0.021\%$ at the four tested
 rates.}
\end{figure}

Let $N$ denote the truncation order,
and $q_{\rm exact}=\epsilon D_{\rm exact}'/D_{\rm exact}$ the logarithmic
derivative of the exact input solution. Define the normalized error
\begin{equation}
 E_N=\frac{|q_{\rm exact}(s_{\rm obs})-
 \sum_{n=0}^{N}\epsilon^nq_n(s_{\rm obs})|}{2\Omega(s_{\rm obs})}.
\end{equation}
Select $N_\ast$ from a five-point upper envelope of $|\epsilon^nq_n|$ to
suppress interference zeros, then evaluate $E_{N_\ast}$ independently.
A weak backward component changes $q$ by magnitude $2\Omega|\beta|$ at
late time, explaining the reflection-sized error in
Fig.~\ref{fig:optimalTruncation}, up to finite-time WKB corrections.
The reference ODE resolves this error, not the much smaller least terms.
The tanh ramp provides an exactly solvable benchmark; the
\href{../../outputs/time_reflection_wkb_coefficients.csv}{coefficient data}
and \href{../../outputs/time_reflection_optimal_truncation.csv}{truncation data}
retain the full numerical checks.

\section{Stable single-mode numerics}

We integrate the unexpanded system~\eqref{eq:BogoliubovODE} using
DOP853 (eighth-order Dormand--Prince), with input $\alpha=1,\beta=0$.
Width scans use $[-18,18]$, relative/absolute tolerances
$2\times10^{-11}/2\times10^{-13}$ and maximum step $\epsilon/10$.
Weak-response and filtered-mode tests use the same window,
$2\times10^{-12}/2\times10^{-14}$ and $\epsilon/12$; the independent original
oscillator uses $\epsilon/15$. Shrinking-feature tests use $[-10,10]$,
the width-scan tolerances and $\epsilon/2$; tighter checks extend to
$[-12,12]$, with $3\times10^{-13}/3\times10^{-15}$ and $\epsilon/3$.
The variation includes the accumulated phase and feedback into $\alpha$.
All coefficients are expressed in the asymptotic phases of
Eq.~\eqref{eq:referenceModes} before subtraction.
At the selected design, changing the half-window from 16 to 20 changes
$|\beta|$ by less than $1.8\times10^{-14}$ absolutely. Independent original-
oscillator integration agrees within $5.4\times10^{-12}$ in complex relative
difference, and the three first-variation implementations agree within
$3.2\times10^{-13}$. The convergence data retain these checks.

\section{Derivative spectrum and finite drive bandwidth}

\begin{figure}[!htbp]
 \centering
 \includegraphics[width=0.90\textwidth]{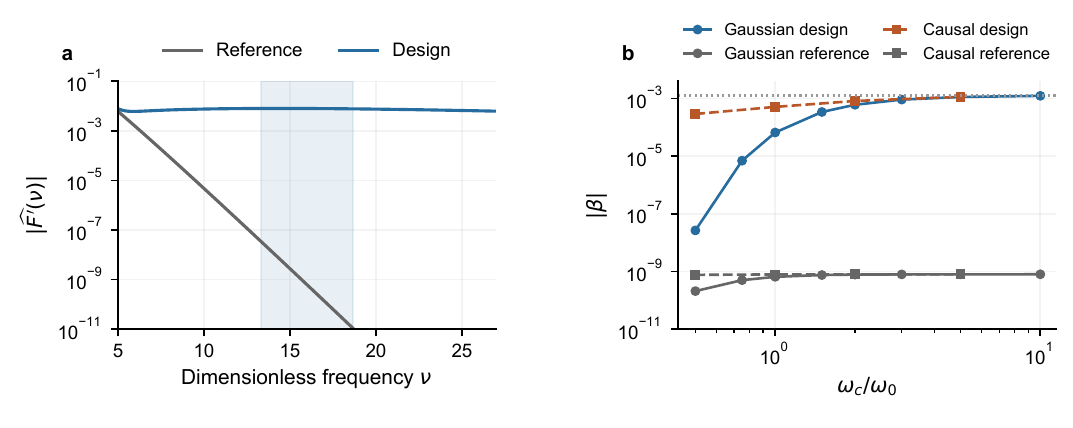}
 \caption{\label{fig:filterBandwidth}Spectral origin and bandwidth cost of
 the optimized reflection. (a) Magnitudes of the Fourier transforms of
 $F'(s)$; the shaded interval is the positive--negative-frequency mixing
 band for $\epsilon=0.15$. (b) Exact single-mode amplitudes after Gaussian
 and causal filtering of each complete ramp. Colored curves show the design,
 gray curves the corresponding references; the horizontal dotted line marks
 the unfiltered design. The response tails control how much of the added
 reflection survives at the same inverse response time $\omega_c$.}
\end{figure}

Freezing the background at $\Omega_{\rm mid}$ in
Eq.~\eqref{eq:betaPerturbationIntegral} gives a local spectral estimate,
with dimensionless mixing frequency $\nu_{\rm mid}=2\Omega_{\rm mid}/\epsilon$:
\begin{equation}
 |\beta_\eta-\beta_0|\simeq
 \frac{\eta\Delta_\Omega^2}{4\Omega_{\rm mid}^2}
 |\widehat{h_a'}(\nu_{\rm mid})|,\qquad
 \widehat{h_a'}(\nu)=\int_{-\infty}^{\infty}
 h_a'(s)e^{-i\nu s}\,ds.
 \label{eq:spectralApprox}
\end{equation}
The phase accumulation rate varies during switching:
$2\Omega(s)/\epsilon$ spans $13.33$--$18.67$ at $\epsilon=0.15$.
This band lies in the exponentially decaying tail of the reference
derivative spectrum. The weak feature adds spectral weight there
[Fig.~\ref{fig:filterBandwidth}(a)]. Physical derivatives and frequencies
are $F'/T$ and $\nu/T$. Quantitative amplitudes below use full mode
integration, not the frozen-frequency estimate.

For an ideal smoothing model, convolve the \emph{entire} prescribed
target ramp $F$ with the normalized Gaussian
$G_\sigma(s)=e^{-s^2/(2\sigma^2)}/(\sqrt{2\pi}\sigma)$:
\begin{equation}
 F_\sigma=G_\sigma*F,\qquad
 \widehat{G_\sigma}(\nu)=e^{-\sigma^2\nu^2/2},\qquad
 \frac{\omega_c}{\omega_0}=\frac{\epsilon}{\sigma}.
 \label{eq:GaussianFilter}
\end{equation}
Here $*$ denotes convolution, $F_\sigma$ the smoothed ramp,
$\tau_b$ the physical response time, $\sigma=\tau_b/T$, and
$\omega_c=1/\tau_b$. This symmetric kernel models ideal smoothing.
Filtering acts on $F$, which is linear in frequency squared. Applying a
voltage or material response requires the corresponding mapping to $F$.

We also use a positive causal kernel and its equivalent drive equation,
\begin{equation}
 K_\sigma(s)=\frac{\Theta(s)}{\sigma}e^{-s/\sigma},\qquad
 \sigma F_{\rm actual}'+F_{\rm actual}=F_{\rm target},\qquad
 \widehat K_\sigma(\nu)=\frac{1}{1+i\sigma\nu}.
 \label{eq:causalResponse}
\end{equation}
Here $\Theta$ is the unit step, $F_{\rm target}$ the prescribed ramp, and
$F_{\rm actual}=K_\sigma*F_{\rm target}$ the realized ramp. At $\omega_c$,
the causal and Gaussian response magnitudes are $1/\sqrt2$ and $e^{-1/2}$.
Both positive kernels preserve endpoints and monotonicity.

We apply each response to both complete ramps, sampled on $[-18,18]$
with spacing $0.0025$. We then use cubic-spline interpolation and integrate
Eq.~\eqref{eq:BogoliubovODE}.
The causal drive equation uses relative/absolute tolerances
$2\times10^{-12}/2\times10^{-15}$ and maximum step $0.01$ in $s$.
Halving the sampling step changes the design amplitudes by less than
$2\times10^{-7}$ relatively at Gaussian cutoffs $\omega_c/\omega_0=1,2$;
at causal cutoff 1, both ramp amplitudes change by less than $3.4\times10^{-6}$.
At $\omega_c=\omega_0$, the Gaussian and causal design amplitudes are
$6.66\times10^{-5}$ and $5.15\times10^{-4}$; Gaussian smoothing at
$0.5\omega_0$ reduces the design amplitude to $2.67\times10^{-8}$.
The surviving reflection therefore depends on the response tail even at
equal response time [Fig.~\ref{fig:filterBandwidth}(b)]. These response
models omit dispersion, loss and waveform noise.

\section{Full-wave Yee-lattice calculation}

We evolve a finite complex analytic signal on a one-dimensional staggered Yee
lattice. Let $j,n$ index position and time, with step sizes $\Delta z,\Delta t$
and $t_n=t_i+n\Delta t$ measured from the initial time $t_i$.
Half indices indicate staggering. In units $c=\varepsilon_0=\mu_0=1$,
the transverse scalar fields obey
\begin{align}
 B_{j+1/2}^{n+1/2}&=B_{j+1/2}^{n-1/2}
 -\frac{\Delta t}{\Delta z}(E_{j+1}^n-E_j^n),\nonumber\\
 D_j^{n+1}&=D_j^n-\frac{\Delta t}{\Delta z}
 (B_{j+1/2}^{n+1/2}-B_{j-1/2}^{n+1/2}),\nonumber\\
 E_j^n&=D_j^n/\varepsilon_r(t_n).
\end{align}
The initial packet is
\begin{equation}
 D(z,t_i)=e^{-(z-z_0)^2/(2\sigma_z^2)}e^{ik_0(z-z_0)}.
\end{equation}
Here $z$ is position, $z_0=60\lambda_0$ the packet center,
$\sigma_z=8\lambda_0$ its envelope width, and $\lambda_0=2\pi/k_0$
the initial carrier wavelength. The domain is $160\lambda_0$ long with
periodic spatial boundaries.
Propagation starts at $s=-L$ and continues along the complete analytic ramp
until $s=L+28\lambda_0/(cT)$, when we project the output.
For the default $L=12$ and $\epsilon=0.15$, this is
approximately $s\in[-12,38.389]$. The additional waiting time separates the
packets; the medium is not frozen at $s=L$.

The magnetic field is initialized mode by mode with the actual velocity
$v_i=1/\sqrt{\varepsilon_r(t_i)}$ at the finite starting time. For constant
wave speed $v$, the discrete angular frequency $\omega_d$ satisfies
\begin{equation}
 \sin\frac{\omega_d\Delta t}{2}=v\frac{\Delta t}{\Delta z}
 \sin\frac{k\Delta z}{2}.
\end{equation}
After switching, projection onto the two discrete eigenvectors gives
\begin{equation}
 |\beta_k|=\sqrt{\frac{\Omega_f}{\Omega_i}}
 \left|\frac{D_{k,-}^{\rm out}}{D_k^{\rm in}}\right|.
\end{equation}
Here $D_k^{\rm in}$ and $D_{k,-}^{\rm out}$ are the incident and outgoing
negative-frequency Fourier amplitudes. The continuum-mode reflected energy
fraction is $(\Omega_f/\Omega_i)|\beta_k|^2$; the packet result averages this
with the incident spectral energy weights. The fine grid has 80 points per
wavelength (ppw); 60- and 40-point
grids test refinement. The central-mode differences from the continuum
result decrease approximately quadratically with grid spacing.
The time step satisfies $v_{\max}\Delta t/\Delta z\leq0.64$, where
$v_{\max}=1.4$ is the maximum wave speed in these units.
We retain 23 positive-$k$ components with incident weight
above $2\times10^{-3}$ of the maximum.

\begin{table}[!htbp]
 \caption{\label{tab:FDTDchecks}Numerical checks supporting the backward packet.
 Grid errors are central-mode relative differences from the continuum
 single-mode amplitude $1.280549\times10^{-3}$. The last two rows report maxima over
 the 23 retained components: the Yee--CT difference is relative, whereas
 the unmodulated floor is $|\beta|$.}
 \begin{ruledtabular}
 \begin{tabular}{lc}
 quantity & value\\
 \hline
 central grid error (80 ppw) & $2.60\times10^{-4}$\\
 central grid error (60 ppw) & $4.62\times10^{-4}$\\
 central grid error (40 ppw) & $1.04\times10^{-3}$\\
 maximum Yee--complex-time difference & $3.08\times10^{-3}$\\
 maximum unmodulated floor & $1.84\times10^{-14}$
 \end{tabular}
 \end{ruledtabular}
\end{table}

Matching the finite-time input at $L=12$ recovers the tanh reference amplitude
$8.05\times10^{-10}$, within $0.42\%$ of the exact result; extending to
$L=14$ changes it by $0.016\%$. The reflected packet-energy fraction is
$2.2965\times10^{-6}$.